\documentclass[10pt]{article}

\usepackage[bottom=1.2in, top=1.2in, right=1.2in, left=1.2in, columnsep=0.81cm] {geometry}

\usepackage[english]{babel}

\usepackage{graphicx}
\usepackage{caption}

\usepackage{amssymb, amsmath, yfonts}
\usepackage{mathrsfs}

\usepackage{listings}
\usepackage{mathtools}
\usepackage{commath}
\usepackage{enumitem}
\usepackage[left]{lineno}
\usepackage{setspace}

\newcommand{\HyFlow}{\textsc{HyFlow}}
\newcommand{\HyFlowPi}{$\pi$\textsc{HyFlow}}

\newcommand{\HyReals}{\mathbb{H}}
\newcommand{\HyRealsp}{\mathbb{H}_0^{+\infty}}
\newcommand{\nullvalue}{\varnothing}

\newcommand{\Xc}{X^\mathrm{c}}
\newcommand{\Yc}{Y^\mathrm{c}}
\newcommand{\Xd}{X^\mathrm{d}}
\newcommand{\Yd}{Y^\mathrm{d}}
\newcommand{\xc}{x_\mathrm{c}}

\newcommand{\xd}{x_\mathrm{d}}

\newcommand{\Lambdac}[1]{\Lambda_{#1}^\mathrm{c}}
\newcommand{\Lambdad}[1]{\Lambda_{#1}^\mathrm{d}}
\newcommand{\lambdad}[1]{\lambda_{#1}^\mathrm{d}}

\newcommand\tab[1][0.35cm]{\hspace*{#1}}
\newcommand{\itemt}{\item\tab}
\newcommand{\itemtt}{\item\tab\tab}
\newcommand{\itemttt}{\item\tab\tab\tab}

\DeclarePairedDelimiter\seqv{\langle}{\rangle}
\DeclarePairedDelimiter\seq{(}{)}

\DeclareMathOperator*{\Cross}{\bigtimes}

\DeclareMathOperator{\while}{while\:}
\DeclareMathOperator{\return}{return\:}

\DeclareMathOperator{\IF}{if\:}

\DeclareMathOperator{\fall}{forall\:}
\DeclareMathOperator{\head}{head}
\DeclareMathOperator{\DO}{do\:}

\newcommand{\powerset}[1]{\mathcal{P}(#1)}

\newcommand{\base}[1]{M_{#1}\:=\:(X,\:Y,\:P,\:P_0,\:\zeta,\:\Pi,\:\pi,\allowbreak\:\sigma,\:\{\Lambda_p\})}
\newcommand{\exec}[1]{M_{#1}\:=(X,\:Y,\:P,\:P_0,\:\zeta,\:\Pi,\:\pi,\allowbreak\:\{\Lambda_p\},\allowbreak\:\Sigma^*,\:\gamma)}
\newcommand{\proc}[2]{M_{#2}^{#1}\:=\:(Y,\:I,\:P,\:P_0,\:\kappa,\:\{\rho_i\},\:\{\omega_i\}, \:\{\varkappa_i\},\:\{\delta_i\},\:\{\Lambdac{i}\},\:\{\lambdad{i}\})}

\newcommand{\quotes}[1]{$``$#1$"$}

\title{$\pi$HyFlow Operational Semantics}
\author{
	Fernando Barros\footnote{Author email address: barros@dei.uc.pt}\\
	University of Coimbra\\Department of Informatics Engineering\\Coimbra, Portugal
}
\date{}
\begin{document}\maketitle

\begin{abstract}

\noindent
Simulation models have been described using different perspectives, or worldviews. In the process interaction world view (PI), every entity is modeled by a sequence of actions describing its life cycle, offering a comprehensive model that groups the events involving each entity. In this paper we describe \HyFlowPi, a formalism for representing hybrid models using a set of communicating processes. This set is dynamic, enabling processes to be created and destroyed at runtime. Processes are encapsulated into \HyFlowPi\ base models and communicate through shared memory. \HyFlowPi, however, can guarantee modularity by enforcing that models can only communicate by input and output interfaces. \HyFlowPi\ extends current PI approaches by providing support for \HyFlow\ concepts of sampling and dense (continuous) outputs, in addition to the more traditional event-based communication. In this paper we present \HyFlowPi\ operational semantics using the concepts of simulator and component.
\end{abstract}

\subsubsection*{Keywords}
Modeling \& simulation, process interaction worldview, hybrid models, operational semantics, co-simulation.

\section{Introduction}

The {\em process interaction worldview} (PI) enables a simple and intuitive description of simulation models. Contrarily to the event interaction approach that offers an unstructured perspective of the systems based on a set of events, commonly represented by event graphs \cite{Schruben-1983:EventGraphs}, the PI organizes events by entity and by their order of occurrence. The result is a script that is easier to understand and verify than the corresponding event graphs. The PI has its origins on the SIMULA language \cite{Gordon-1978:GPSS}. Some simulation languages supporting PI favors the active client approach \cite{Henriksen-198I:GPSS}. In this view, transitory entities, like clients, are represented by processes, while permanence entities, like servers, are represented as data structures. In the alternative active server, processes model the permanent resources of the systems, like machines, while clients are viewed as passive data that is passed among processes. This view is often considered a requirement for a modular representation of systems, and it is supported by formalisms like DEVS \cite{Zeigler-1976:TMS} for describing discrete event systems, and \HyFlow\ \cite{Barros-2017:Chattering}, to represent hybrid systems. Although modularity leverages hierarchical models, and the ability to represent complex models by a composition of simpler ones, it does not always provide the most adequate level of representation for simple models. In fact, base models in these types of modular representations can only describe one event, forcing, in general, that models with two or more events to be represented as a composition of several base models, one for each event. Considering, as an example, a system with one queue feeding two servers; a single event model needs to decompose the system into three models and describe the synchronization among entities. On the contrary, in PI, the communication can be achieved, in a simple way, through shared memory data structures since simulation processes have non-preemptive semantics. However, simulation languages supporting PI often do not enforce modularity, making it difficult to represent complex systems.

In this paper we present a new formalism to represent hierarchical, modular hybrid models that enables, at the base level, the ability to represent several events, keeping the advantages of PI. Our goal is to combine the simplicity of PI for describing small systems, with hierarchical, and modular constructs that enable a systems-of-systems representation for revealing system behavior and components’ interactions \cite{Nielson-2015:SoS}. Modularity also makes easier to represent systems with a dynamic topology, by providing a one-to-one mapping between changes in system topology and the corresponding structural adaptation of the model. We also show that the ability to create processes in runtime enables PI to support the active client approach, while guaranteeing modular and hierarchical models.

In previous work, we have developed the Hybrid Flow System Specification (\HyFlow), a formalism aimed to represent hierarchical and modular hybrid systems \cite{Barros-2017:Chattering}. \HyFlow\ defines sampling and the exact representation of continuous signals as first-order constructs, enabling a simple specification of pull-communication in addition to push-communication, typical of discrete event systems. \HyFlow\ models exhibit a dynamic topology, making it possible to make arbitrary changes in model composition and coupling. In this paper we develop the \HyFlowPi\ formalism, an extension of the \HyFlow\ formalism, to represent base hybrid models using the process interaction worldview. \HyFlowPi\ can describe simulation processes that can communicate through both sampling and discrete events. The operational semantics of the \HyFlowPi\ formalism is provided using the concepts of simulator and component.

\section{The \HyFlowPi\ Formalism}\label{section:HyFlowPi}

\begin{figure}[hptb]
	\centering
	\includegraphics[width=0.5\textwidth] {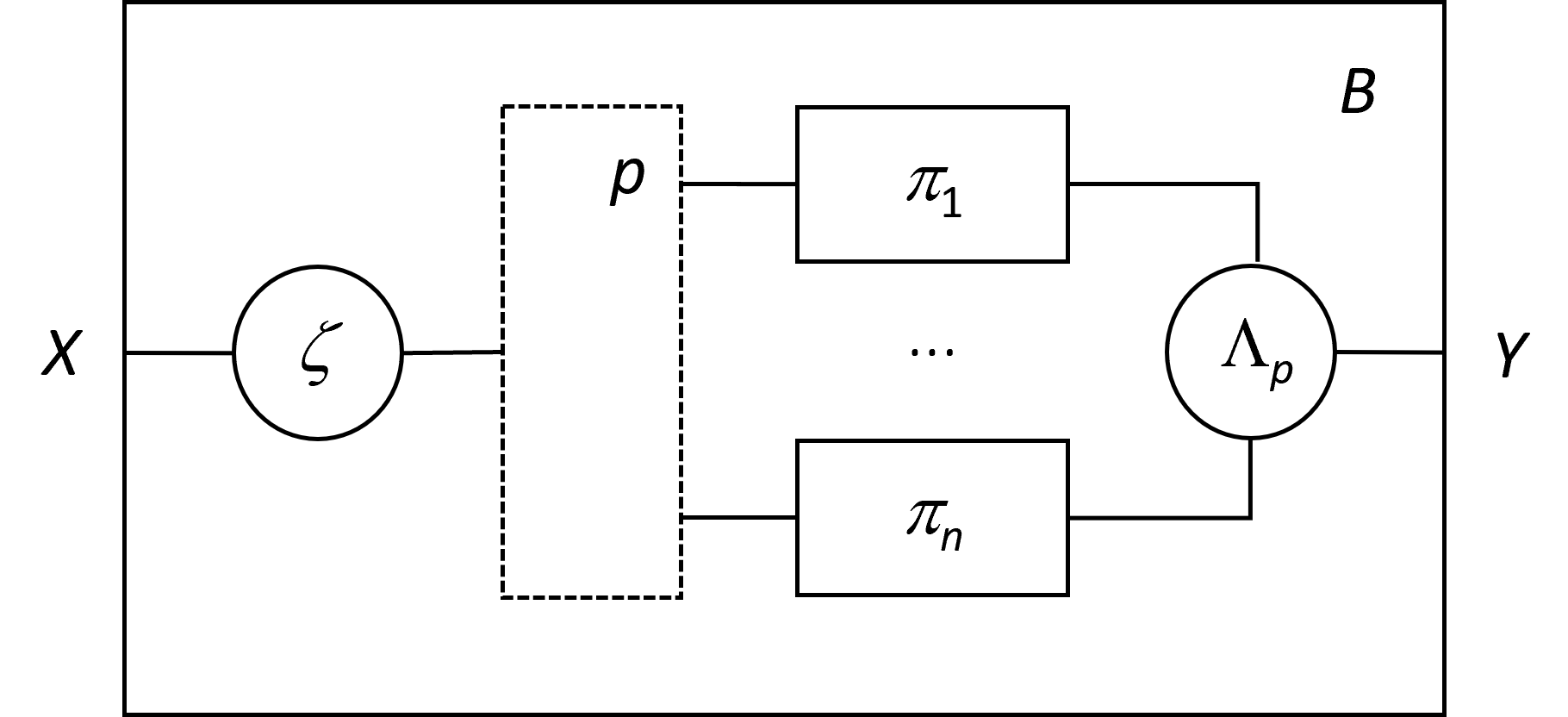}
	\caption{\HyFlowPi\ base model structure.}
	\label{fig:rcwv}
\end{figure}

A \HyFlowPi\ base model defines a modular entity that encloses a set of processes $\{\pi_1, ..., \pi_n\}$ communicating through a shared p-state $p$, as represented in Figure \ref{fig:rcwv}. Each process keeps its own (private) p-state, and it can perform read/write operations on the shared p-state. Processes have suspend/resume semantics but are non-preemptive making them implementable by coroutines.
Base models communicate through modular input ($X$) and output ($Y$) interfaces. Processes do not have an input, since after their creation they are suspended at some point on their flows waiting for being reactivated. Process communication with the external entities can only thus be achieved, indirectly, through the shared p-state $p$, that can be modified by all processes and by the model input function $\zeta$. Processes, however, have a modular output interface. Base model output is computed by function $\Lambda = \{\Lambda_p\}$ that uses the outputs from all processes. Since \HyFlowPi\ base models have a modular interface, they can be composed to form networks as described in Section \ref{section:network_model}. A formal definition of the \HyFlowPi\ base model is provided in the next section.

\subsection{ \HyFlowPi\ Base Model}\label{section:base_model}

A \HyFlowPi\ base model defines a modular entity enclosing a set of processes that communicate through a shared p-state. Each process keeps its own (private) p-state and can perform read/write operations on the shared p-state. Processes are non-preemptive making them implementable by coroutines. While threads have long been available in most programming languages, the native support for coroutines in the C++ high performance language is recent, being introduced by C++20 (ISO/IEC 14882 standard). Formally, a \HyFlowPi\ base model associated with name $B$ is defined by:
\[\base{B},\] where:
\begin{description}\setlength\itemsep{0pt}
	\item $X = \Xc \times \Xd$ is the set of input flow values, with
	\itemt $\Xc$ is the set of continuous input flow values,
	\itemt $\Xd$ is the set of discrete input flow values, and
	\itemt $X^\nullvalue = \Xc \times (\Xd + \nullvalue)$,
	\item $Y = \Yc \times \Yd$ is the set of output flow values, with
	\itemt $\Yc$ is the set of continuous output flow values,
	\itemt $\Yd$ is the set of discrete output flow values, and
	\itemt $Y^\nullvalue = \Yc \times (\Yd+ \nullvalue)$,
	\item $P$ is the set of partial shared states (p-states),
	\item $P_0$ is the set of (valid) initial p-states,
	\item $\zeta: P \times X^\nullvalue \longrightarrow P$ is the input function,
	\item $\Pi$ is a set of processes names,
	\item $\pi: P \longrightarrow \powerset{\Pi}$ is the current-processes function, where $\mathcal{P}$ is the power set operator,
	\item $\sigma: \powerset{\Pi} \longrightarrow \mathcal{P}^*(\Pi)$ is the ranking function, where $\mathcal{P}^*(\Pi)$ is the set of all sequences based on set $\Pi$, constrained to: $\sigma(C|\: C \subseteq \Pi) =\seq{c_1, \ldots, c_n} \Rightarrow \{c_1, \ldots, c_n\} = C \wedge |\sigma(C)| = |C|$,
	\item for all $p \in P:$
	\itemt $\Lambda_p: \Cross\limits_{i\:\in\:(\sigma \circ \pi)(p)} Y_i^\nullvalue \longrightarrow Y^\nullvalue$ is the output function associated with p-state $p$.
\end{description}

\noindent
Since processes have no entry points, the representation of model input needs to be stored in the shared p-state so it can be accessed by the processes. The input function $\zeta$ is responsible for updating the current p-state when the model receives a value either through sampling or event communication. The set of processes is dynamic, being the current set given by function $\pi$. Given processes can access the shared p-state only one can be active at any time. The ranking function $\sigma$ decides process resume order. The output function $\lambda = \{\lambda_p\}$ maps the outputs of all processes into the output associated with the base model. In the next section we provide the formal description of a \HyFlowPi\ process.

\subsection{\HyFlowPi\ Process Model} \label{section:process_model}

A process is a sequence of actions that usually take some amount of virtual (simulation) time to be executed. Processes are coordinated by the base model. The base model chooses a process that can be executed and resumes it. After executing, the process suspends itself and gives the control back to base model. This explicit invocation is necessary since simulation processes are non-preemptive. Given a base model $M_B = (X, Y_B, P_B, P_{0, B}, \zeta, \Pi, \pi, \sigma, \{\Lambda_p\})$, the model of a process $\varpi \in \Pi$ is defined by:
\[\proc{B}{\varpi},\] where:
\begin{description}\setlength\itemsep{0pt}
	\item $Y$ is the set of output flow values,
	\itemt $\Yc$ is the set of continuous output flow values
	\itemt $\Yd$ is the set of discrete output flow values
	\item $I$ is the set of indexes,
	\item $P$ is the set of p-states,
	\item $P_0$ is the set of (valid) initial p-states,
	\item $\kappa: P \longrightarrow I$ is the index function,
	\item for all $i \in I:$
	\itemt $\rho_i: P \longrightarrow \HyRealsp$ is the time-to-input function,
	\itemt $\omega_i: P \longrightarrow \HyRealsp$ is the time-to-output function,
	\itemt $\varkappa_i: P \times P_B \longrightarrow \{\top, \bot\}$ is the condition function,
	\itemt $\delta_i: S \times P_B \longrightarrow P \times P_B$ is the transition function,
	\itemt $\Lambdac{i}: S \times P_B \longrightarrow \Yc$ is the continuous output function,
	\itemt $\lambdad{i}: P \times P_B \longrightarrow \Yd$ is the partial discrete output function,
	\itemt $\Lambdad{i}: S \times P_B \longrightarrow \Yd \cup \{\nullvalue\}$ is the discrete output function defined by:
	\itemtt $\Lambdad{i}((p, e), p_B) =
		\begin{cases*}
			\lambdad{i}(p, p_B) & if $(e = \omega_i(p))$\\
			\nullvalue & otherwise
		\end{cases*}$
	\itemt with $S = \{(p,\:e)|\:p \in P,\:0 \leq e \leq \nu_{\kappa(p)}(p)\}$, the state set,
	\itemt and $\nu_i(p) = \min\{\rho_i(p), \omega_i(p)\}, i = \kappa(p)$, is the time-to-transition function,
\end{description}

\noindent
For time specification, \HyFlowPi\ uses the set of hyperreal numbers $\HyReals$, that enables to express causality, by assuming that a transition occurring at time $t$, changes process p-state at time $t + \varepsilon$, where $\varepsilon \in \HyReals$ is an infinitesimal.

A process defines only its output, while the {\em input} is {\em inferred}, as mentioned before, from base model p-state. A process defines its own (private) p-state, for reducing inter process dependency. A process dynamic behavior is ruled by six structured/split functions, being the segments currently active determined by the index function $\kappa$. The active function segments associated with p-state $p \in P$ are $(\rho_i, \omega_i, \varkappa_i, \delta_i, \Lambdac{i}, \lambdad{i})|_{i = \kappa(p)}$.

The time-to-input-function $\{\rho_i\}$ specifies the interval for sampling (reading) a value. Since each process specifies its own reading interval, sampling is made asynchronously, and it can be made independently by any process. The time-to-output-function $\{\omega_i\}$ specifies the interval to produce (write) a discrete flow. The condition function $\{\varkappa_i\}$, checks whether the process has conditions to run given base model p-state and its own p-state. While $\{\rho_i\}$ and $\{\omega_i\}$ specify a time interval for process re-activation, $\{\varkappa_i\}$ checks if the process can be re-activated at the {\em current time}. Function $\{\delta_i\}$ specifies process and base model p-states after process re-activation. Function $\{\Lambdac{i}\}$ specifies process continuous output flow, and $\{\Lambdad{i}\}$ specifies process discrete output flow. The former can be non-null at every time instant, while the latter can only be non-null at a finite number of time instants during a finite time interval. We note that $\{\Lambdac{i}\}$ can provide an exact description for an arbitrary continuous signal based on a discrete formalism. Some approaches are limited to piecewise constant representations of continuous signals \cite{Bastian-2011:Master}, \cite{Lee-2005:Operational}.

A simulation process gets most of its modeling expressiveness by enabling the dynamic switching of the active functions that describe its behavior. The semantics of base models and processes are given in the next sections.

\subsection{\HyFlowPi\ Base Component} \label{section:base_component}

The semantics of \HyFlowPi\ base models is described using the concept of {\em component}. The alternative concept of {\em iterative system specification} \cite{Barros-2002:CFSS} could also be used. We found, however, that components can express model semantics in a simpler form, when using the set of hyperreals, instead the set of real numbers, to define time. Additionally, the component concept provides a simple description of dynamic network models. Given that base model semantics rely on process semantics, we refer here informally to the latter, postponing the formal description to the next section.

For each base model there is an associated component that is responsible for its simulation. The component provides base model operational semantics enabling \HyFlowPi\ base model unambiguous interpretation and implementation. A component is the actual entity that is placed into simulation. As such, a component keeps its current state and defines a set of actions to compute its next state, and component output, for example. Actions are described using base model definition.

 In component definition a variable $v$ is represented by $\seqv{v}$. In action definition, the assignment to variable $v$ of a value $x$ is represented by $v \gets x$, and The \quotes{forall} operator represents an iteration over a set or sequence. An action can also behave like a function and {\em return} a value (using the keyword \quotes{return}). A base component associated with model $\base{B}$, is defined by:
\[\mathcal{C}_B = (\seqv{v, p}, N, \Omega, \Delta),\]
where:\vspace{-0pt}

\begin{linenumbers}\resetlinenumber\setlength\linenumbersep{-2pt}\modulolinenumbers[2]
\begin{description}\setlength\itemsep{2pt}
	\item $v \in Y^\nullvalue$, is the output value variable,
	\item $p \in P$, is the current p-state variable,
	\item $N\dblcolon\:\longrightarrow \HyReals$, is the component next time transition action, defined by:
	\begin{description}\setlength\itemsep{2pt}
		\item $\mathrm{N}()\triangleq \return \min\limits_{i\:\in\:\pi(p)}\{\mathrm{N}_i()\}$
	\end{description}
	\item $\Omega \dblcolon\ \HyReals$, is the output action, defined by:
	\begin{description}\setlength\itemsep{2pt}
		\item $\Omega(t) \triangleq v \gets \Lambda_p(\Cross\limits_{i\:\in\:(\sigma\circ\pi)(p)} \Lambda_i(t, p))$,
	\end{description}
	\item $V\dblcolon\longrightarrow Y^\nullvalue$, is the output value action, defined by:
	\begin{description}\setlength\itemsep{2pt}
		\item $V() \triangleq \return v$
	\end{description}

	\item$\Delta\dblcolon\HyReals \times X^{\nullvalue}$, is the transition action, defined by:
	\begin{description}\setlength\itemsep{2pt}
		\item $\Delta(t,\:(\xc,\:\xd)) \triangleq$
		\itemt $\IF (t \neq \mathrm{N}() \land \xd = \nullvalue) \return$
		\itemt $p \gets \zeta(p, (\xc,\:\xd))$
		\itemt $M \gets \{i|\:i \in \pi(p) \wedge N_i() = t\}$
		\itemt $\fall(i \in \sigma(M))\:p \gets \Delta_i(p)$
		\itemt $\DO \{$
		\itemtt $Z \gets \{i |\:i \in \pi(p) \wedge K_i(p))\}$
		\itemtt $\IF (Z \neq \phi)\:\{$
		\itemttt $i \gets \head(\sigma(Z))$
		\itemttt $p \gets \Delta_i(t,\:p)$
		\itemttt $M \gets M + i$
		\itemtt $\}$
		\itemt $\} \while (Z \neq \phi)$
		\itemt $\fall (i \in M)\:U_i(t)$
	\end{description}
\end{description}
\end{linenumbers}

\noindent
A base component associated with model $M_B$ keeps its output in variable $v$ (line 1). The current p-state $p$ is stored in line 2. Next time transition action $N$ (line 3), computes the minimum re-activation time (next time) of the inner/child processes (line 4). Base component output action (line 5) sets variable $v$ according to the outputs of the inner processes (line 6). Variable $v$ value can be accessed through the output value action (lines 7-8). The transition action (line 9) defines base model next state based on the current time and the input value. The condition of line 11 simplifies network component transition definition (described in Section \ref{section:network_component}). Line 12 computes component the new p-state based on the input value. Line 13 finds the set of processes that are scheduled to undergo a transition at time $t$. These processes are ranked and triggered at time $t$, possibly changing the base component p-state (line 14). Since the p-state has changed, we find (line 16), the processes that can be re-activated, i.e., the processes whose conditions functions evaluate to $\top$. Line 18 chooses the process with the highest rank, which is re-activated in line 19. This sequence is repeated until there is no more process that can undergo a conditional transition (line 22). A process can be triggered several times in this cycle, since the guard condition also depends on the shared p-state that, in general, is modified at each transition. All processes that have undergone a transition will execute an update action (line 23). This action is described in the next section.

\subsection{\HyFlowPi\ Process Simulator}\label{section:process_simulator}

The process simulator is responsible for updating process p-state according to the corresponding model. A process simulator describes the coroutine-like (non-preemptive) semantics of each \HyFlowPi\ process. For processes we use the simulator concept since, contrarily to base components, a simulator is a non-modular entity that can only exist within a specific parent base component. In a practical implementation, coroutines representing processes can only exist within some context, an object for example, being able to share memory with the coroutines in the same context. We emphasize that a \HyFlowPi\ process provides just a set of operators letting process semantics largely undefined and open to different interpretations. The simulator is required to precisely define process dynamic-behavior/operational-semantics. A simulator keeps the output value, the current p-state, and the time when the last transition has occurred. A simulator for process $\varpi$ with model $\proc{B}{\varpi}$ and associated with base model $\base{B}_B$ is defined by:
\[\mathcal{S}_\varpi^B= (\seqv{v, (p, t_L)}, N, \Omega, V, K, \Delta, U),\] where:\vspace{2pt}

\begin{linenumbers}\resetlinenumber\setlength\linenumbersep{0pt}\modulolinenumbers[2]
\begin{description}\setlength\itemsep{2pt}
	\item $v \in Y^\nullvalue$, is the output value,
	\item $(p,\: t_L)$ is the simulator current state, with $p \in P$ the simulator p-state and $t_L \in \HyRealsp$ the transition time to current p-state $p$,
	\item $N\dblcolon\:\longrightarrow \HyReals$, is the next transition time, defined by:
	\begin{description}\setlength\itemsep{2pt}
		\item $\mathrm{N}()\triangleq \return t_L + \min\{\rho_{\kappa(p)}(p), \omega_{\kappa(p)}(p)\}$
	\end{description}
	\item $\Omega\dblcolon\HyReals \times P_B$, is the output action, defined by:
	\begin{description}\setlength\itemsep{2pt}
		\item $\Omega(t, p_B) \triangleq$
		\itemt $e \gets t - t_L$
		\itemt $v \gets (\Lambdac{\kappa(p)}((p,\: e), p_B), \Lambdad{\kappa(p)}((p,\: e), p_B))$
	\end{description}
	\item $V\dblcolon\longrightarrow Y^\nullvalue$, is the output value, defined by:
	\begin{description}\setlength\itemsep{2pt}
		\item $V() \triangleq \return v$
	\end{description}
	\item $K\dblcolon P_B \longrightarrow \{\top,\: \bot\}$, is the condition verification, defined by:
	\begin{description}\setlength\itemsep{2pt}
		\item $K(q) \triangleq \return \varkappa_{\kappa(p)}(p,\:q)$
	\end{description}
	\item$\Delta\dblcolon\HyReals \times P_B \longrightarrow P_B$, is the transition action, defined by:
	\begin{description}\setlength\itemsep{2pt}
		\item $\Delta(t,\:q) \triangleq$
		\itemt $(p,\: q') \gets \delta_{\kappa(p)}((p,\:t - t_L),\: q)$
		\itemt $\return q'$
	\end{description}
	\item $U\dblcolon\HyReals $, is the time update defined by:
	\begin{description}\setlength\itemsep{2pt}
		\item $U(t) \triangleq t_L \gets t + \varepsilon$
	\end{description}
\end{description}
\end{linenumbers}
\noindent
Lines 4-5 define simulator next transition time. The output action stores the current output value in variable $v$ (lines 6-9). This value can be read by action $V$ (lines 10-11). The condition verification action (lines 12-13) checks whether the process has conditions to run. The transition action (lines 14-17) computes the new simulator and parent base component p-states. The time update action (line 18) sets process simulator $t_L$ (line 19).

\subsection{\HyFlowPi\ Network Model}\label{section:network_model}

\HyFlowPi\ networks are composed by base or other network models. Additionally, each network has a special component, named as the executive, that is responsible for defining network topology (composition and coupling). A \HyFlowPi\ network model associated with name $N$ is defined by:
\[ M_N = (X, Y, \eta),\] where:
\begin{description}\setlength\itemsep{2pt}
	\item $X = \Xc \times \Xd$ is the set of network input flows,
	\item $Y = \Yc \times \Yd$ is the set of network output flows,
	\item $\eta$ is the name of the dynamic topology network executive.
\end{description}
\noindent
The executive model is a \HyFlowPi\ base model extended with topology related operators. This model is defined by:
\[\exec{\eta},\]
\begin{description}\setlength\itemsep{2pt}
	\item where:
	\itemt $\Sigma^*$ is the set of network topologies,
	\itemt $\gamma: P \longrightarrow \Sigma^*$ is the topology function.
\end{description}\vspace{2pt}
\noindent
The network topology $\gamma(p_\alpha) \in \Sigma$, corresponding to the p-state $p_\alpha \in P$, is given by:
\begin{eqnarray*}
		\gamma(p_\alpha) =
		(C_\alpha, \{I_{i, \alpha}\} \cup \{I_{\eta, \alpha}, I_{N, \alpha}\}, \{F_{i, \alpha}\} \cup \{F_{\eta, \alpha}, F_{N, \alpha}\}),
\end{eqnarray*}
where:\vspace{2pt}
\begin{description}\setlength\itemsep{0pt}
	\item $C_\alpha$ is the set of names associated with the executive p-state $p_\alpha$,
	\item for all $i \in C_\alpha + \eta$:
	\item\tab$I_{i, \alpha}$ is the sequence of influencers of $i$,
	\item\tab$F_{i, \alpha}$ is the input function of $i$,
	\item $I_{N, \alpha}$ is the sequence of network influencers,
	\item $F_{N, \alpha}$ is the network output function,
\end{description}
\begin{description}\setlength\itemsep{2pt}
	\item for all $i \in C_\alpha$
	\itemt $M_i = (X, Y, P, P_0, \zeta, \Pi, \pi, \sigma, \{\Lambda_p\})$, for base models,
	\itemt $M_i = (X, Y, \eta)$, for network models.
\end{description}
\noindent
Variables are subjected to the following constraints for all $p_\alpha \in P_\alpha$:

\begin{linenumbers}\resetlinenumber\setlength\linenumbersep{0pt}\modulolinenumbers[2]
\begin{description}\vspace{0pt}
	\item $N \not\in C_\alpha$,
	\item $\eta \not\in C_\alpha$,
	\item $N \not\in I_{N,\alpha}$,
	\item $F_{N,\alpha}: \Cross\limits_{k\:\in\:I_{N,\alpha}} Y_k \longrightarrow Y^\nullvalue$,
	\item $F_{i,\alpha}: \Cross\limits_{k\:\in\:I_{i,\alpha}} V_k \longrightarrow X_i^\nullvalue$,
	\begin{singlespace}\vspace{2pt}
		\itemt where $V_ k =
		\begin{cases}
			Y_k^\nullvalue & \IF k \neq N\\
			X^\nullvalue & \IF k = N
		\end{cases}$
	\end{singlespace}
	\item $F_{N,\alpha}((v_{\mathrm{c}, k_1}, \nullvalue), (v_{\mathrm{c}, k_2}, \nullvalue), ... ) = (y_{\mathrm{c}, N}, \nullvalue)$,
	\item $F_{i, \alpha}((v_{\mathrm{c}, k_1}, \nullvalue), (v_{\mathrm{c}, k_2}, \nullvalue), ... ) = (x_{\mathrm{c}, i}, \nullvalue)$.
\end{description}
\end{linenumbers}
Constraints 1 and 2 impose that the executive cannot remove neither the network nor itself. Constraint 3 enforces causality (Section \ref{section:process_model}). Constraints 7 and 8 are a characteristic of discrete systems and impose that a non-null discrete flow cannot be created from a sequence composed exclusively of null discrete flows.

The topology of a network is defined by its executive through the topology function $\gamma$, which maps the executive p-state into network composition and coupling. Topology adaption can thus be achieved by changing executive p-state. A \HyFlowPi\ network model is simulated by a \HyFlowPi\ network component that performs the orchestration of network inner components. Network simulation is achieved by a general communication protocol that relies only on the component interface. This protocol is independent from model details, enabling the composition of components that are handled as black boxes.

\subsection{\HyFlowPi\ Executive Component}\label{section:executive_component}

Before describing the network component, we define first the executive component, an extension to the base component, that introduces the topology function required to establish network topology. A \HyFlow\ executive component $\mathcal{C}_\eta$ associated with the executive model $\exec{\eta}$ is defined by:
\[ \mathcal{C}_\eta = (\seqv{v, (p, t_L)}, N, \Omega, \Delta, \Gamma),\] where: \vspace{-0pt}
\begin{description}\setlength\itemsep{2pt}
	\item$\Gamma\dblcolon\:\longrightarrow \Sigma^*$, is the executive topology action defined by:
	\itemt$\Gamma() \triangleq \return \gamma(p)$
\end{description}
\vspace{ -0pt}
\noindent
The $\Gamma$ action returns network topology at the current time, based on executive current p-state. Since the executive model is an extension of the base model, the executive component inherits the actions previously defined for the base component.

\subsection{\HyFlowPi\ Network Component}\label{section:network_component}

A network component is composed by one executive component and a set of other components. As mentioned before, components and their interconnections can change according to executive p-state. An exception being the executive that cannot be removed. Components can be base or other \HyFlowPi\ network components, making it possible to define networks hierarchically. A \HyFlowPi\ network component $\Xi_N$ associated with the network model $M_N = (X, Y, \eta)$, executive $\exec{_\eta}$, and current topology $\Gamma_\eta() = (C, \{I_i \} \cup \{I_\eta, I_N\}, \{F_i\} \cup \{F_\eta, F_N\})$, is defined by:
\[\mathcal{C}_N = (\seqv{v}, N, \Omega, \Delta),\] where:

\begin{linenumbers}\resetlinenumber\setlength\linenumbersep{0pt}\modulolinenumbers[2]
\begin{description}\setlength\itemsep{2pt}
	\item $v \in Y$, is network component output,
	\item $\mathrm{N}\dblcolon\longrightarrow \HyReals$, is the next transition time, defined by:
	\itemt $\mathrm{N}()\triangleq \return \min \{\mathrm{N}_i()|\:i\:\in\:C + \eta\}$
	\item $\Omega\dblcolon\HyReals$, is the network output action, defined by:
	\itemt $\Omega(t)\triangleq v \gets F_N(\Cross\limits_{i\: \in\: I_N} \Lambda_i(t))$

	\item $V\dblcolon \longrightarrow Y^{\nullvalue}$, is the network output value action, defined by:
	\itemt $V()\triangleq \return v$

	\item $\Delta\dblcolon\HyReals \times X^{\nullvalue}$, is the network component transition, defined by:
	\itemt $\Delta(t,\:x)\triangleq$
	\itemtt forall $(i \in C)\:\Delta_i(t, F_i(\Cross\limits_{j\:\in\: I_i} v_j))$
	\itemtt $\Delta_\eta(t, F_i(\Cross\limits_{j\:\in\: I_i} v_j))$
	\itemttt with $v_j =
		\begin{cases}
			\mathrm{V}_j() & \IF j \neq N\\
			x & \IF j = N
		\end{cases}$
\end{description}
\end{linenumbers}
\noindent
Network next transition time (lines 2-3) is defined as the minimum transition time of network components. Network output is defined based on the output of its inner components (lines 4-5). This value is stored in line 1 and can be accessed by the output action (lines 6-7). The transition action (line 8) is divided in two steps. In the first step, components, except the executive, compute their own transition (line 10). To simplify transition description, the decision to actually make the transition is made by each component, as mentioned in Section \ref{section:base_component}. The executive transition is only performed as the last one (line 11), since the new topology will only be used after transition time, at $t + \epsilon$.

\subsection{\HyFlowPi\ Component Simulation}\label{section:coordinator}

To perform a simulation, it is necessary to define a mechanism to execute component transitions according to their time advance specification. The simulation of a component, both base or network, is performed by the action:
\[\mathcal{S}\dblcolon\mathfrak{C}\times\HyReals,\]
where $\mathfrak{C} = \{c|\:M_c = (\{\} \times \Xd, Y, \ldots)\}$, is the set of names associated with \HyFlowPi\ models (base or network) defining a null continuous input flow interface. The simulation action is defined by:\vspace{-0pt}

\begin{linenumbers}\resetlinenumber\setlength\linenumbersep{0pt}\modulolinenumbers[2]
\begin{description}\setlength\itemsep{2pt}
	\item $\mathcal{S}(c,\:end)\triangleq$
	\itemt $clock \gets N_c()$
	\itemt $\while(clock < end)\:\{$
	\itemtt $\Omega_c(clock)$
	\itemtt $\Delta_c(clock, (\nullvalue,\:\nullvalue))$
	\itemtt $clock \gets N_c()$
	\itemt $\}$
\end{description}
\end{linenumbers}\resetlinenumber
\noindent
The simulation loop involves a sequence of steps: compute current component output (line 4), trigger component transition at time $clock$ (line 5) and compute the time of next transition that becomes the new $clock$ (line 6). Line 4 enforces causality, since it computes component output before the transition is performed.

\section{Related Work}\label{section:related}

The concepts of continuous flow and generalized sampling were introduced in the Continuous Flow System Specification (CFSS) formalism \cite{Barros-2002:CFSS}. CFSS enables the exact representation of continuous signals in digital computers. These signals can be read using non-uniform sampling. Different components can also sample signals asynchronously. The support for multiple clocks was introduced in the Esterel language \cite{Berry-2001:Esterel}, but continuous flows were limited to piecewise constant segments \cite{Berry-2001:Esterel}.

The Hybrid Flow System Specification (\HyFlow) formalism \cite{Barros-2017:Chattering} combines continuous flows \cite{Barros-2002:CFSS} and discrete events \cite{Zeigler-1976:TMS}. The process interaction worldview (PI) was introduced by SIMULA \cite{Gordon-1978:GPSS}. The formal specification of PI was introduced in \cite{Zeigler-1976:TMS}. This work, however, was limited to discrete event non-modular models, based on a static set of processes. This approach also provides a limited view on process conditional waiting. A more general approach to this problem was proposed in \cite{Cota-1992:PI}. This work, however, did not add any support to model modularity.

Traditional representations of ODEs rely on the analog computer paradigm and require no explicit representation of numerical methods \cite{Henzinger-1996:HybridAutomata}. In these approaches, the modeler only needs to describe the ODEs \cite{Praehofer-1991:Combined}, and in some cases to choose the numerical method for handling the overall set of ODEs \cite{Fritzson-2003:Modelica}. Although this seems a good solution, since it frees users from numerical details, it has several limitations. The co-simulation \cite{Bastian-2011:Master} is not guaranteed since ODEs need to be transformed and converted into a set of first-order ODEs and collectively solved by a single numerical integrator. Another limitation of these approaches is the difficulty to introduce new numerical integrators since they are not explicitly represented. Given that these frameworks only offer ODEs as first-class constructs, solutions requiring the combination of different families of numerical integrators can also not be described.

\section{Conclusion}\label{section:conclusion}

The \HyFlowPi\ formalism provides a representation for hierarchical, and modular hybrid systems. This formalism is intended to support modeling \& simulation for performance evaluation. \HyFlowPi\ uses the concepts of continuous flow and generalized sampling to describe continuous systems.
\HyFlowPi\ networks have a time-varying topology leveraging a simple description of systems that undergo structural changes. \HyFlowPi\ introduces the support for processes into the original \HyFlow\ formalism. \HyFlowPi\ ability to dynamically create/destroy processes provides a framework to combine the active client and the active server process interaction worldviews.


\begin{thebibliography}{10}
	
	\bibitem{Barros-2002:CFSS}
	F.~Barros.
	\newblock Towards a theory of continuous flow models.
	\newblock {\em International Journal of General Systems}, 31(1):29--39, 2002.
	
	\bibitem{Barros-2017:Chattering}
	F.~Barros.
	\newblock Chattering avoidance in hybrid simulation models: A modular approach
	based on the {HyFlow} formalism.
	\newblock In {\em Symposium on Theory of Modeling and Simulation}, 2017.
	
	\bibitem{Bastian-2011:Master}
	J.~Bastian, C.~Clau\ss, S.~Wolf, and P.~Schneider.
	\newblock Master for co-simulation using {FMI}.
	\newblock In {\em Proceedings of the 8th Modelica Conference}, 2011.
	
	\bibitem{Berry-2001:Esterel}
	G.~Berry and E.~Sentovich.
	\newblock Multiclock {E}sterel.
	\newblock In {\em Correct Hardware Design and Verification Methods}, volume
	2144 of {\em LNCS}, pages 110--125, 2001.
	
	\bibitem{Cota-1992:PI}
	B.~Cota and R.~Sargent.
	\newblock A modification of the process interaction world view.
	\newblock {\em ACM Transactions on Modeling and Computer Simulation},
	2(2):109--129, 1992.
	
	\bibitem{Fritzson-2003:Modelica}
	P.~Fritzson.
	\newblock {\em Principles of Object-Oriented Modeling and Simulation with
		{Modelica} 2.1}.
	\newblock Wiley, 2003.
	
	\bibitem{Gordon-1978:GPSS}
	G.~Gordon.
	\newblock {\em {System Simulation}}.
	\newblock Prentice-Hall, Englewood Cliffs, New Jersey, 2nd edition, 1978.
	
	\bibitem{Henriksen-198I:GPSS}
	J.~Henriksen.
	\newblock {GPSS} - finding the apropriate world-view.
	\newblock In {\em Winter Simulation Conference}, pages 505--516, 1981.
	
	\bibitem{Henzinger-1996:HybridAutomata}
	T.~Henzinger.
	\newblock The theory of hybrid automata.
	\newblock In {\em 11th Annual IEEE Symposium on Logic in Computer Science},
	pages 278--292, 1996.
	
	\bibitem{Lee-2005:Operational}
	E.~Lee and H.~Zheng.
	\newblock Operational semantics of hybrid systems.
	\newblock In {\em Hybrid Systems Computation and Control}, volume 3414 of {\em
		LNCS}, pages 392--406, 2005.
	
	\bibitem{Nielson-2015:SoS}
	C.~Nielson, P.~Larsen, J.~Fitzgerald, J.~Woodcock, and J.~Peleska.
	\newblock Systems of systems engineering: Basic concepts, model-based
	techniques, and research directions.
	\newblock {\em ACM Computing Surveys}, 48(2):1--41, 2015.
	
	\bibitem{Praehofer-1991:Combined}
	H.~Praehofer.
	\newblock Systems theoretic formalisms for combined discrete-continuous system
	simulation.
	\newblock {\em International Journal of General Systems}, 19(3):219--240, 1991.
	
	\bibitem{Schruben-1983:EventGraphs}
	L.~Schruben.
	\newblock Simulation modeling with event graphs.
	\newblock {\em Communications of the ACM}, 26(11):957--963, 1983.
	
	\bibitem{Zeigler-1976:TMS}
	B.~Zeigler.
	\newblock {\em Theory of Modelling and Simulation}.
	\newblock Wiley, 1976.
	
\end{thebibliography}

\end{document}